\documentstyle[12pt]{article}
\begin{document}

\title{HARTREE-FOCK-BOGOLIUBOV \\ FOR DEFORMED NEUTRON-RICH NUCLEI}

\author{
  NAOKI TAJIMA\\
  Department of Basic Science,
  University of Tokyo, Komaba,\\ 153-8902, Japan\\
  E-mail: tajima@nt1.c.u-tokyo.ac.jp
}

\date{}

\maketitle

\begin{abstract}
The two-basis method to solve the HFB for deformed nuclei in
coordinate space is examined concerning the precision of the density
tail.  Small cutoff energies are shown to give rise to ripples in the
tail, whose wave length corresponds to the cutoff momentum.  More
precise solutions require higher cutoff energies, which result in HFB
matrices of larger dimensions.  To circumvent the difficulty of the
large dimension, we employ another method to solve the HFB ---the
natural-orbital method introduced originally for spherical nuclei---
to apply it to deformed nuclei. The method can be successfully
implemented with a three-dimensional Cartesian mesh representation.
\end{abstract}

\section{Introduction} \label{sec:intro}


In neutron-rich nuclei, the effects of continuum states on the pairing
correlation are expected to play an important role.
As for nuclei near the neutron drip line, it is obvious that the
continuum states are strongly involved in the pairing correlation
because the Fermi level $\lambda_{\rm n}$ and the pairing gap
$\Delta_{\rm n}$ have similar sizes $\sim \pm 1$ MeV.  The effect of
such a continuum-state pairing may be so strong that the neutron drip
line can be pushed outward by several nuclei.\cite{DNW96}

For nuclei with $\lambda_{\rm n} \stackrel{>}{\sim} -5$ MeV, too, the
continuum states should be taken into account explicitly because the
pairing-active space for HFB calculations should be larger than one
major shell, i.e., 
$\epsilon_{\rm cut}$ $>$ $\lambda$ $+$ $\hbar \omega$, 
in order to take into account a correct size of shell effects.


The pairing correlation has been treated usually in the BCS
approximation, which relies on an assumption that the pair-scattering
matrix elements $\langle \psi_i \psi_{\bar{\imath}} | V_{\rm p} |
\psi_j \psi_{\bar{\jmath}} \rangle$ do not depend on the form of the
wavefunctions $\psi_i$ and $\psi_j$, e.g. a constant.  This assumption
results in a situation that the nucleus is surrounded by unphysical
dilute neutron gas when $\epsilon_{\rm cut} > 0$.
Therefore, to include the positive energy states in the pairing
correlation, one needs to switch from the HF(Hartree-Fock)+BCS to the
Hartree-Fock-Bogoliubov (HFB) scheme in coordinate space.


So far, several methods have been presented to solve the HFB equation.
For spherical nuclei, there are three methods:\\
1) Radial differential equations (Dobaczewski et al., 1984),\cite{DFT84}\\
2) Finite-element method for finite-range pairing forces
(Ring et al., 1997),\cite{PVL97}\\
3) Natural-orbital representation (Reinhard et al, 1997).\cite{RBR97}

For deformed nuclei, two methods have been presented. Although
spherical cases can be solved easily with present computers, deformed
cases in coordinate space are still a challenge.\\
4) Diagonalization in the harmonic-oscillator basis 
(Gogny et al., 1980),\cite{DG80}\\
5) Two-basis method (Heenen et al., 1994).\cite{GBD94,THF96,TFH97}

In this paper, we describe the formulations in the subsequent two
sections and then study two subjects: First, in
section~\ref{sect:tail}, we examine the two-basis method concerning
the precision of the low-density tail at large radius as a function of
the cutoff energy, i.e., the number of HF orbitals included in the
diagonalization.  We show that more precise solutions require higher
cutoff energies, which result in HFB matrices of larger dimensions.
Second, in section~\ref{sect:implhfbn}, we test the natural-orbital
method for deformed cases in a three-dimensional (3D) Cartesian mesh
representation.
  
\section{HFB in coordinate space}

In this section, we formulate the HF and the HFB in the
coordinate-space representation in order to elucidate a difficulty of
the HFB and to suggest its possible solution in terms of the
natural-orbital representation.  For the sake of simplicity, in this
section, we consider only one kind of nucleons and designate the
number of nucleons by $N$.  The spin of a nucleon is represented by
$s$.

\subsection{HF} \label{sect:hf}

In the HF, one should minimize $\langle \Psi | H | \Psi \rangle$
for single Slater-determinant states,
\begin{eqnarray}
|\Psi \rangle & = & \prod_{i=1}^{N} a^{\dagger}_i | 0 \rangle, \\
a^{\dagger}_i & = & \sum_{s} \int d \vec{r} \; \psi_{i} (\vec{r},s) \;
a^{\dagger}_{\vec{r}s} 
\;\;\; \mbox{ : single-particle state},
\end{eqnarray}
by varying $\{ \psi_i \} _{i=1,\cdots,N}$ under orthonormality
conditions $\langle \psi_i | \psi_j \rangle$ = $\delta_{ij}$.

\subsection{HFB} \label{sect:hfb}

In the HFB, the state takes the following form,
\begin{eqnarray}
|\Psi \rangle & = & \prod_{i=1}^{\# {\rm basis}} b_i | 0 \rangle, 
\label{eq:hfbpsi}\\
b_i & = & \sum_{s} \int d \vec{r} \left\{ 
\phi_i^{\ast} (\vec{r},s) \; a_{\vec{r}s} +
\varphi_i (\vec{r},s) \; a^{\dagger}_{\vec{r}s} \right\} 
\;\;\; \mbox{ : quasi-particle state}, \label{eq:hfbqp}
\end{eqnarray}
where ``\#basis'' is the number of basis states of the employed
representation: For a 3D-mesh representation,\cite{BFH85} it is the
number of the mesh points (times four when spin-orbit potentials are
included) and typically $10^4$-$10^5$.  One should vary 
$\{ \phi_i, \varphi_i \}_{i=1,\cdots,\# {\rm basis}}$ 
under appropriate orthonormality conditions.

The essential difference between the HF and the HFB is that one has to
consider only $N$ $\sim 10^2$ wavefunctions in the former while one
has to treat explicitly as many single-particle wavefunctions as the
number of the basis in the latter.

\subsection{HFB with the two-basis method}  \label{sect:twobasis}

In this method, the HFB equation is solved by
diagonalizing the HFB matrix in a single-particle basis
$\{ \psi_i \}_{i=1,\cdots,K}$ 
consisting of the eigenstates of the mean-field
hamiltonian $h$ (excluding the pairing potential): 
$h \psi_i$ = $\epsilon_i \psi_i$.  
The number of the basis, $K$, is determined by a cutoff energy 
$\epsilon_{\rm cut}$ as
$\epsilon_1$ $\le \cdots \le$ $\epsilon_K$ $ \le$
$\epsilon_{\rm cut}$ $\le$ $\epsilon_{K+1}$ $\le \cdots$.  We will show in
section~\ref{sect:tail} that $K \gg N$ to obtain high-precision
density tails.

%
\begin{figure}[bth]           
\begin{center}
\framebox[4cm]{Figure \ref{fig:hfbpict}}
\end{center}
\caption{Schematic picture to explain the different number of
single-particle states necessary to express the HFB ground state between the
two-basis and the natural-orbital methods.
\label{fig:hfbpict}}
\end{figure}
%

\subsection{HFB in natural-orbital basis}  \label{sect:hfbn}

\noindent
Owing to the Bloch-Messiah theorem,
the state (\ref{eq:hfbpsi}) can be expressed as,
\begin{eqnarray}
|\Psi \rangle & = & \prod_{i=1}^{\# {\rm basis}} 
\left( u_i + v_i \; a^{\dagger}_{i} \; a^{\dagger}_{\bar{\imath}} \right)
| 0 \rangle, \label{eq:hfbnpsi}\\
a^{\dagger}_i & = & \sum_{s} \int d \vec{r} \; \psi_{i} (\vec{r},s) \;
a^{\dagger}_{\vec{r}s} 
\;\;\; \mbox{ : natural orbital (canonical basis)}. \label{eq:hfbna}
\end{eqnarray}
One should vary $\{ \psi_i , u_i \}_{i=1,\cdots,\# {\rm basis}}$
under constraints for orthonormality,
\begin{equation} \label{eq:orthon}
     \langle \psi_i | \psi_j \rangle = \delta_{ij} \;\;\;  
     ( 1 \leq i \leq j \leq \mbox{{\#}basis}),
\end{equation}
and for the expectation value of the number of nucleons, 
$2 \sum_{i=1}^{K} v_i^2 = N$.

Reinhard et al.\ regard the advantage of the representation
(\ref{eq:hfbnpsi}) over (\ref{eq:hfbpsi}) to be that one has to
consider only a single set of wavefunctions $\{ \psi_i \}$ unlike a
double set $\{\phi_i, \varphi_i \}$.\cite{RBR97} However, we expect
more benefit from the natural-orbital representation. Namely, $i$ may
be truncated as $i \leq K$ = ${\cal O}(N)$ $\ll$ $\# {\rm basis}$ to a
very good approximation.  It is because $\psi_i$ should be a localized
function while the orthogonality does not allow many wavefunctions to
exist in the vicinity of the nucleus.  This situation is illustrated
in Fig.~\ref{fig:hfbpict}.
For 3D-mesh representations, \# basis is proportional to the volume of
the cavity (box) while $K$ is proportional to the volume of the
nucleus. The latter is $10^1$-$10^2$ times as small as the former.

\section{HFB with density-dependent zero-range forces} \label{sect:hfbddd}

\subsection{Interaction} \label{sect:int}

We employ density-dependent zero-range interactions in the rest of this paper
for the sake of simplicity. There will not be essential differences
in the formulation if we use the full-form Skyrme force.
The force is expressed in the parameterization of the Skyrme force as,
\begin{equation} \label{eq:mfint}
V(\vec{r}_1,s_1;\vec{r}_2,s_2) = \left( t_0 + \frac{1}{6} t_3 
\rho\left( \vec{r}_1 \right)^{\alpha} \right) 
\delta \left( \vec{r}_1 - \vec{r}_2 \right).
\end{equation}
We adopt
$t_0$ = $-1099$ MeV fm$^{3}$,
$t_3$ = $17624$ MeV fm$^{3+3 \alpha}$, and $\alpha$ = 0.98 
(it is not 1 only for the sake of a test of the code) when the force
is used to construct the mean-field (HF) potential.\cite{Koo76} 
We use a different strength to make the pairing potential. We express
the force in the parameterization of Ref.\cite{TBF93}:
\begin{equation} \label{eq:pairint}
V_{\rm p}(\vec{r}_1,s_1;\vec{r}_2,s_2)  = v_{\rm p} 
\frac{1-P_{\sigma}}{2}
\left( 1 - \frac{\rho (\vec{r}_1)}{\rho_{\rm c}}
\right) 
\delta \left( \vec{r}_1 - \vec{r}_2 \right).
\end{equation}
We use $\rho_{\rm c}$ = 0.32 fm$^{-3}$ 
(to roughly vanish the volume-changing effect\cite{TBF93}).

\subsection{Hamiltonian density} \label{sect:hamdens}

For the sake of simplicity, we treat $N$=$Z$ nuclei without
Coulomb interaction in the rest of this paper. 
Then, the state of the nucleus is expressed as, 
\begin{equation} \label{eq:totwf}
| \Psi \rangle  = \prod_{i=1}^{K}
\left( u_i + v_i \; a^{\dagger}_{i} \; a^{\dagger}_{\bar{\imath}} 
\right)_{\rm proton}
\left( u_i + v_i \; a^{\dagger}_{i} \; a^{\dagger}_{\bar{\imath}} 
\right)_{\rm neutron}
| 0 \rangle.
\end{equation}
With the interactions (\ref{eq:mfint}) and (\ref{eq:pairint}), the
total energy for state (\ref{eq:totwf}) is written as,
\begin{eqnarray}
E & = & \langle \Psi | H | \Psi \rangle = 
\int {\cal H}\left(\vec{r}\right) d \vec{r}, \\
{\cal H} & = &
\frac{\hbar^2}{2m} \tau + \frac{3}{8}t_0 \rho^2 +\frac{1}{16}t_3
\rho^{2+\alpha} 
+\frac{1}{8}v_{\rm p} 
\left( 1 - \frac{\rho}{\rho_{\rm c}} \right) \tilde{\rho}^2,
\end{eqnarray}
where $\tau$ is the kinetic energy density,
%
%
\begin{equation}
\rho (\vec{r}) =  4 \sum_{i=1}^{K} v_i^2 | \psi_i (\vec{r}) |^2,
\;\;\;\;\;
\tilde{\rho}(\vec{r})  =  4 \sum_{i=1}^{K} u_i v_i | \psi_i (\vec{r}) |^2.
\end{equation}
The stationary condition
$\delta {\cal H} = 0 $ leads to a mean-filed and a pairing potentials:
\begin{eqnarray}
h & = -\frac{\delta E}{\delta \rho} & = 
\frac{\hbar^2}{2m} \vec{\nabla}^2 + \frac{3}{4}t_0 \rho +\frac{2+\alpha}{16}
t_3 \rho^{1+\alpha} -\frac{v_{\rm p}}{8 \rho_{\rm c}} \tilde{\rho}^2, \\
\tilde{h} & = \frac{\delta E}{\delta \tilde{\rho}} & = 
\frac{1}{4} v_{\rm p} 
\left( 1 - \frac{\rho}{\rho_{\rm c}} \right) \tilde{\rho}.
\end{eqnarray}

\section{Test of the accuracy of the two-basis method for HFB}
\label{sect:tail}

We have developed from scratch an HFB program for spherical cases
based on the two-basis method explained in sect.~\ref{sect:twobasis}.
It is used to examine the precision of the low-density tail at large
radius as a function of the cutoff energy $\epsilon_{\rm cut}$.

In Fig.~\ref{fig:ripples} we show the
density profiles for $(Z,N)=(38,68)$
calculated with the Braghin-Vautherin force\cite{BV94} 
and $v_{\rm p}$ = $-400$ MeV.
One can see that
(i) The density is localized with accuracy $\sim 10^{-6}$ fm$^{-3}$.
(ii) The cutoff makes ripples in the tail.  The wave-length of the
ripples agrees with $2\pi\hbar(2m \epsilon_{\rm cut})^{-1/2}$, i.e.,
half of the de~Broglie wavelength for $\epsilon_{\rm cut}$.
%
%

\begin{figure}[bth]           
\begin{center}
\framebox[4cm]{Figure \ref{fig:ripples}}
\end{center}
\caption{Density profiles of HF and two-basis HFB solutions.
\label{fig:ripples}}
\end{figure}

To obtain a more precise density tail, one has to increase
$\epsilon_{\rm cut}$, which results in an increase of $K$, the number
of single-particle wavefunctions to be considered explicitly.  $K$
increases only slowly as $\epsilon_{\rm cut}^{1/2}$ for spherical case
for each angular momentum, while it grows rapidly as $\epsilon_{\rm
cut}^{3/2}$ for deformed case.  The rapid grow of the latter case
makes the two-basis method practically inapplicable to deformed nuclei
because the bottle-necks in computation of mean-field methods on 3D
meshes are the parts which spend computation time proportional to
$K^2$ like the orthogonalization. Therefore, one need a different
method to solve the HFB for deformed nuclei with enough high cutoff
energies.

\section{Implementation of the natural-orbital HFB method on 3D mesh}
\label{sect:implhfbn}

The natural-orbital method explained in sect.~\ref{sect:hfbn} was
originally introduced for spherical nuclei.\cite{RBR97} We have
implemented the method to treat deformed nuclei in a 3D-mesh
representation.  

First, let us present a summary of the formulation, which has some
differences from Ref.\cite{RBR97} Instead of minimizing $E$ = $\langle
\Psi | H | \Psi \rangle$ with $|\Psi\rangle$ given by
Eq.~(\ref{eq:totwf}) under constraints of Eq.~(\ref{eq:orthon}), one
may introduced a Routhian $R$,
\begin{equation} \label{eq:routhian}
R = E - \epsilon_{\rm Fermi} \cdot 4 \sum_{i=1}^{K} v_i^2 -
\sum_{i=1}^{K} \sum_{j=1}^{K} \lambda_{ij} \left\{ 
\langle \psi_i | \psi_j \rangle - \delta_{ij} \right\}, \;\;
\lambda_{ij} = \lambda_{ji}^{\ast},
\end{equation}
and minimize it without constraints.  In the above definition, in
order to make $R$ real for the sake of convenience, $K^2$ Lagrange
multipliers $\lambda_{ij}$ obeying hermiticity are introduced instead
of $\frac{1}{2}K(K+1)$ independent multipliers.  Note that
$\delta_{ij}$ is subtracted from $\langle \psi_i | \psi_j \rangle$ in
order to treat $\lambda_{ij}$ not as constants like 
$\epsilon_{\rm Fermi}$ but as functionals of the wavefunctions.

Stationary conditions of $R$ result in the following equations.
\begin{eqnarray}
\frac{\partial R}{\partial v_i}=0 & \Rightarrow &
v_i^2 = \frac{1}{2} - \frac{1}{2} \frac{h_{ii} -\lambda}
{\sqrt{(h_{ii}-\lambda)^2+\tilde{h}_{ii}^2}}, 
\;\;\mbox{assuming} \; \tilde{h}_{ii}\le 0, \label{eq:bcs} \\
\frac{\delta R}{\delta \psi_i^{\ast}}=0 & \Rightarrow &
{\cal H}_i \psi_i - \sum_{j=1}^{K} \lambda_{ij} \psi_j
-\sum_{j=1}^{K} \sum_{k=1}^{K} \frac{\delta \lambda_{jk}}{\delta \psi_i^{\ast}}
\left\{ \langle \psi_j | \psi_k \rangle - \delta_{jk} \right\} = 0,
\label{eq:drdp}\\
& & {\cal H}_i = v_i^2 h + u_i v_i \tilde{h}.\label{eq:hi}
\end{eqnarray}
For HF, the orthogonalization conditions are easily realized because
$\psi_i$ are eigenstates of the same hermite operator $h$ and thus are
orthogonal at the solution: The orthogonalization procedure is needed
only because the orthogonality is unstable.  On the other hand, for
the natural-orbital HFB method, the orthogonalization is essential
because the single-particle hamiltonians ${\cal H}_i$ differs from
state to state.  Therefore, the determination of the explicit
functional form of $\lambda_{ij}$ is the most important part of the
method.  Reinhard et al.\ have proposed
\begin{equation} \label{eq:lambda1}
\lambda_{ij} = \frac{1}{2} 
\langle \psi_j | \left( {\cal H}_i + {\cal H}_j \right) | \psi_i \rangle.
\end{equation}
We can justify their choice as follows:
From the requirement that Eq.~(\ref{eq:drdp}) must hold at the solution
(where $\langle \psi_i | \psi_j \rangle = \delta_{ij}$),
one deduces,
\begin{equation} \label{eq:lambda2}
\lambda_{ij} = \langle \psi_j | {\cal H}_i | \psi_i \rangle
\;\;\; \mbox{at the solution}.
\end{equation}
Eqs.~(\ref{eq:lambda1}) and (\ref{eq:lambda2}) are equivalent at the
solution because $\lambda_{ij}$ is defined to be hermite.  Since this
hermiticity must hold at any points, one should not adopt
Eq.~(\ref{eq:lambda2}) but Eq.~(\ref{eq:lambda1}).  
Because what is needed is Eq.~(\ref{eq:lambda2}) and the hermiticity,
one may use more complex forms like,
\begin{equation} \label{eq:lambda3}
\lambda_{ij}= \frac{1}{2} 
\langle \psi_j | \left( {\cal H}_i + {\cal H}_j \right) | \psi_i \rangle
\left\{ 
    1 + f_2 \left( 
                   \langle \psi_j | {\cal H}_i | \psi_i \rangle -
                   \langle \psi_j | {\cal H}_j | \psi_i \rangle
            \right)^2
\right\},
\end{equation}
where $f_2$ is a parameter to maximize the convergence speed.

To obtain the HFB solutions in the natural-orbital formalism, one
can utilize the gradient method,
which includes the imaginary-time evolution method:
\begin{equation} \label{eq:imagevo}
\psi_i \rightarrow 
\psi_i - \Delta \tau 
\frac{\delta R}{\delta \psi_i^{\ast}}
\end{equation}

We have developed a 3D-mesh natural-orbital HFB program from scratch
according to the above formulation.  First of all, we test the
feasibility of the method for $^{40}$Ca.  The wavefunctions are
expressed with $39 \times 39 \times 39$ mesh points with mesh spacing
of 0.8 fm.  Note that the requirement of precision is higher for
HFB than for HF because one has to treat larger momentum components
than the Fermi momentum in HFB.  We employed the 17-point
finite-difference approximation to the Laplacian.  The vanishing
boundary conditions are imposed on the boundary (0th and 40th
mesh points) and the wavefunctions are anti-symmetrically reflected in the
boundary to apply the finite-difference formula.  We considered
$K=20$ natural orbitals, which can contain 80 (=$2 \times A$)
nucleons.  

\begin{figure}[tbh]           
\begin{center}
\framebox[4cm]{Figure \ref{fig:conv}}
\end{center}
\caption{Convergence of HF and HFB in the natural-orbital method.
\label{fig:conv}}
\end{figure}

We show an example of the convergence history in Fig.~\ref{fig:conv}.
In this calculation, we set $f_2=0$ in Eq.~(\ref{eq:lambda3})
and $\Delta \tau$ = $10^{-24}$ sec in Eq.~(\ref{eq:imagevo}) .
We neglect $\delta \lambda / \delta \psi^{\ast}$ in
Eq.~({\ref{eq:drdp}).  Instead, at every 50 steps, $\{ \psi_i \}$ are
Gram-Schmidt orthogonalized in the ascending order of $h_{ii}$ and
then the HFB hamiltonian is diagonalized in the basis to renew $\{
\psi_i, v_i \}$.

In the left-hand portion, the error of Eq.~(\ref{eq:drdp}), i.e., ${\rm
max}_{i=1,\cdots,K} |{\cal H}_i \psi_i - \sum_j \lambda_{ij} \psi_j |$
are plotted versus the evolution step.  The corresponding quantity for
HF, ${\rm max}_{i=1,\cdots,A/4} |h \psi_i - \langle \psi_i | h |
\psi_i \rangle \psi_i |$ , is also plotted.  The figure demonstrates
that one can indeed obtain HFB solutions with the natural-orbital HFB
method in the 3D-mesh representation.  We obtained similar convergence
curves for the error of the orthogonality and for the inconsistency
between the potential and the densities.  The right-hand portion shows the
error of the total energy (estimated as the difference of the total
energy from the convergent value), which also shows an exponential
convergence pattern.

The speed of the convergence is, however, about ten times as slow as
the HF case.  We are now trying to improve the convergence speed, the
stability of the evolution, and the robustness of the method.

\section{Influence of pairing correlation on deformation}

As examples of the applications of our natural-orbital
HFB program, we show in Fig.~\ref{fig:triax} the change of shape
due to pairing correlation.  The root-mean-square values of
$x$, $y$, and $z$ are plotted as functions of
the strength of the pairing interaction $v_{\rm p}$ for $^{32}$S and
$^{60}$Zn.  Both nuclei are triaxial when the pairing correlation is
off (pairing gap $\Delta =0$). However, as soon as the pairing
correlation sets in, the axial symmetry is restored.

The pairing correlation does not always restore symmetric shapes but
sometimes break a symmetry present without pairing: For $^{60}$Zn, the
expectation value of $\frac{1}{2}(1-{\tau}_z)Y_{30}$ is zero before the
pairing sets in while it is about half the Weiskopf unit when the
pairing correlation is present.

\begin{figure}[tbh]           
\begin{center}
\framebox[4cm]{Figure \ref{fig:triax}}
\end{center}
\caption{Triaxiality of HFB solutions versus the strength of the pairing
interaction.
\label{fig:triax}}
\end{figure}

\section*{Acknowledgments}

The author thanks to Prof.~J.~Dobaczewski, Prof.~P.-H.~Heenen,
Prof.~D.~Brink and Prof.~N.~Onishi for discussions during and after
the symposium.


\vspace*{\baselineskip}
\fbox{ \fbox{ \begin{minipage}[t]{140mm}
This paper has been published in
the proceedings of the XVII RCNP international symposium
on Innovative Computational Methods in Nuclear Many-Body Problems
--Towards a new generation of physics in finite quantum systems--
(INNOCOM97), Osaka, Japan, November 10-15, 1997,
edited by H. Horiuchi, M. Kamimura, H. Toki, F. Fujiwara, M. Matsuo,
and Y. Sakuragi,
(1998) World Scientific (Singapore), pp.~343-351.

The present address of the author is:\\
Department of Applied Physics, Fukui University,\\
Bunkyo 3-9-1, Fukui, 910-8507, Japan\\
E-mail: tajima@quantum.apphy.fukui-u.ac.jp
\end{minipage} } }

\end{document}